\documentclass[10pt,aps,pra,reprint]{revtex4-1}
\usepackage{amsmath,amssymb}
\usepackage[utf8]{inputenc}
\usepackage{graphicx}
\usepackage{color}
\usepackage[english]{babel}
\usepackage[encapsulated]{CJK}

\begin{document}

\begin{CJK*}{UTF8}{}

\title{Kinetic bandgap analysis of plasma photonic crystals}
\author{Jan Trieschmann}
\author{Thomas Mussenbrock}
\affiliation{Electrodynamics and Physical Electronics Group, Brandenburg University of Technology Cottbus--Senftenberg, Siemens-Halske-Ring 14, 03046 Cottbus, Germany}
\date{\today}

\begin{abstract}

The dispersion relation of plasma and plasma-dielectric photonic multilayer structures is approached in terms of a one-dimensional Particle-in-Cell simulation. For several plasma-dielectric configurations, the system response is obtained using a pulsed excitation and a subsequent two-dimensional frequency analysis. It is first shown that the dispersion relation of a single, homogeneous plasma slab is well described by the cold-plasma model even at low pressures of 1 Pa. The study is extended to the simulation of plasma photonic crystals with a variety of configurations, based on the work of Hojo and Mase [J. Plasma Fusion Res. \textbf{80}, 89 (2004)]. Considering a one-dimensional plasma photonic crystal made from alternating layers of dielectric and homogeneous plasma slabs, it is shown that the assumption of a cold-plasma description is well justified also in this case. Moreover, in this work the results are reformatted and analyzed in a band diagram representation, in particular based on the lattice constant $a$. Based on these considerations a scaling invariant representation is presented, utilizing a generalized set of parameters. The study is completed with an exemplary comparison of three plasma-dielectric photonic crystal configurations and their corresponding band diagrams.

\end{abstract}

\maketitle

\end{CJK*}

\newpage

\section{Introduction}
In the field of optics and photonics, photonic crystals (PC) \cite{yablonovitch_inhibited_1987} are well known for their manifold unique properties \cite{saleh_fundamentals_2007}. Many aspects, in particular concerning the possibility to guide electromagnetic fields at low losses have found applications in information technology \cite{russell_photonic_2003}. Moreover, these structures allow for a wide range of scientific as well as industrial applications (e.g., mode confinement and guiding or frequency filters for information processing). Influenced by the unique properties of plasmas in the microwave regime, in the past decade intensive research has been done on merging the properties of dielectric bandgap structures and plasmas \cite{wu_chirped_2005,sakai_plasmas_2012,sakai_functional_2013,wang_plasma_2016,semnani_plasma-enabled_2016,parsons_microwave_2017,qu_properties_2017,gregorio_reconfigurable_2017}. Various aspects of plasma photonic crystals (PPC) have been investigated concerning the theoretical prediction of photonic bandgaps (PBGs) and frequency filtering in one and more dimensions \cite{hojo_dispersion_2004,kong_anomalous_2010}, the theoretical prediction of two-dimensional PPCs and their experimental validation \cite{sakai_verification_2005,sakai_photonic_2007,sakaguchi_photonic_2007,lo_reconfigurable_2010,varault_plasma-based_2011,gregorio_reconfigurable_2017}. In most theoretical models used so far, the plasma has been introduced in terms of the simplest approach available, namely the cold-plasma model. Thus, a simple dispersion relation can be obtained with the assumption of a Maxwell-Boltzmann distribution of electrons. Using a reduced kinetic theory \cite{shkarofsky_particle_1966}, it can be shown that a similar mathematical description can be obtained even for non-Maxwellian electron energy distributions. A more fundamental (i.e., fully kinetic) approach to PPCs has been pursued using Particle-in-Cell (PIC) simulations \cite{yin_bandgap_2009,li-mei_dispersion_2010}. To the best of our knowledge, however, an in-depth study of the plasma dispersion characteristics in the collisional case remains due. The aim of this work is to approach this limitation using PIC simulations.

\section{Simulation setup}
In order to study the interaction of electromagnetic waves with plasma-dielectric structures on a kinetic basis, a one-dimensional PIC code is employed \cite{birdsall_plasma_1991,birdsall_particle--cell_1991,boris_relativistic_1970}. The code used in this work is an extension of the benchmarked \cite{turner_simulation_2013} \textit{yapic}.The code is extended to include a one-dimensional full-wave description of the perpendicular electromagnetic fields $\vec{E}(x,t) = E(x,t) \, \vec{e}_z$ and $\vec{B}(x,t) = -B(x,t) \, \vec{e}_y$. The procedure uses the finite-difference time-domain (FDTD) method \cite{yee_numerical_1966,taflove_computational_2005} with absorbing boundary conditions. The plasma is coupled to the electromagnetic fields using the current density of the moving charged particles similar to the procedure of \cite{villasenor_rigorous_1992,omura_one-dimensional_2005}. The structure of interest is analyzed from the response of the system to a given small-signal electromagnetic excitation. For the electric field, a short modulated pulse of the form
\begin{align*}
	E(x=x_\textrm{c}, t) = E_0 \left[ 0.5 + \cos(\omega_0 t) + \cos(2\omega_0 t) \right] \exp\left[ -(t/\tau_\textrm{p})^2 \right]
\end{align*}
with a pulse width $\tau_\textrm{p} = 8.5$ ps, a fundamental frequency $\omega_0 = 2\pi \cdot 55$ GHz, and a field strength $E_0 = 10^6$ V/m is excited in the center of the configuration at $x_\textrm{c}$. The system response is thereafter evaluated for a duration $t_\textrm{max} = 2.25$~ns. Subsequently, the thus obtained spatio-temporal evolution of the electric field is analyzed in the two-dimensional frequency domain (i.e., spatial and temporal) using the fast Fourier transform (FFT) method. Monte Carlo collisions (MCC) \cite{skullerud_stochastic_1968,mertmann_fine-sorting_2011} are used to simulate a plasma operated at a pressure $p=1$~Pa in argon gas at a temperature $T=300$~K \cite{phelps_application_1994,lxcat_plasma_2012}. The plasma is initialized with an electron temperature of $T_\textrm{e} = 3$ eV and a homogeneous plasma density $n_0 = 10^{13}~\textrm{cm}^{-3}$. At the boundary interfaces reflection boundary conditions are imposed for the particles, while the electromagnetic field is unaffected (i.e., transparent). To reduce numerical heating, a spatial discretization smaller than the Debye length is chosen $\Delta x \ll \lambda_\textrm{D} \approx 4.1~\mu$m. The time step follows from the Courant stability criterion $\Delta t < \Delta x / c$ \cite{taflove_computational_2005}.

In the following, two different configurations are investigated: (i) a homogeneous plasma slab with length $L_\textrm{s} = 33.6$ cm and (ii) a plasma-dielectric photonic crystal (detailed later; cf. figure \ref{fig:schematic}). The configurations are detailed in the respective subsections.

\section{Results}
\subsection{Homogeneous plasma slab}

\begin{figure}[b!]
\centering
\resizebox{8cm}{!}{
\includegraphics{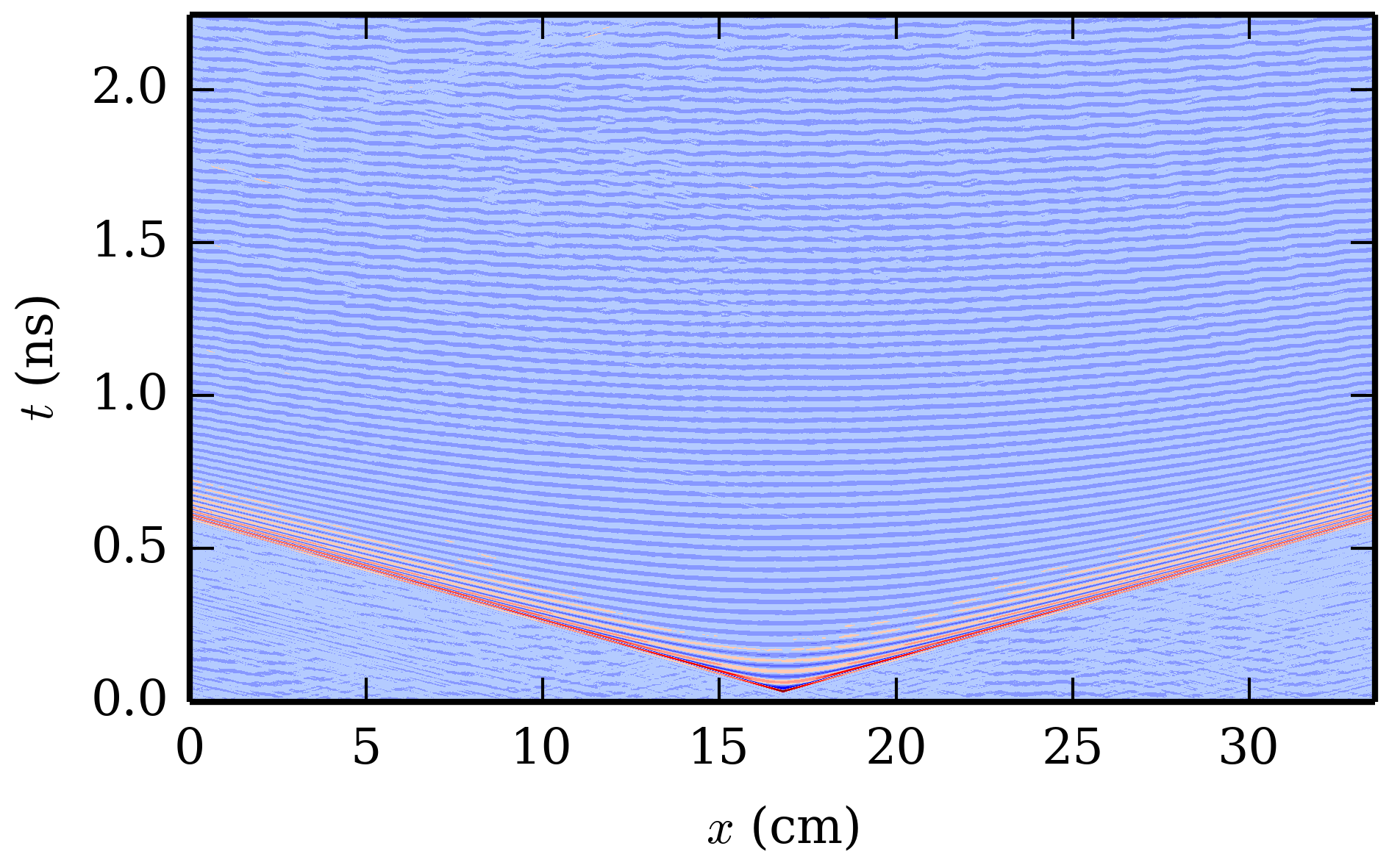}
}
\caption{Spatio-temporal plot of the electric field propagation through the homogeneous plasma slab as calculated using PIC. Color min/max from blue to red are $[-2,~2.4] \cdot 10^6$ V/m.}
\label{fig:plasma_propagation}
\end{figure}

\begin{figure}[t!]
\centering
\resizebox{8cm}{!}{
\includegraphics{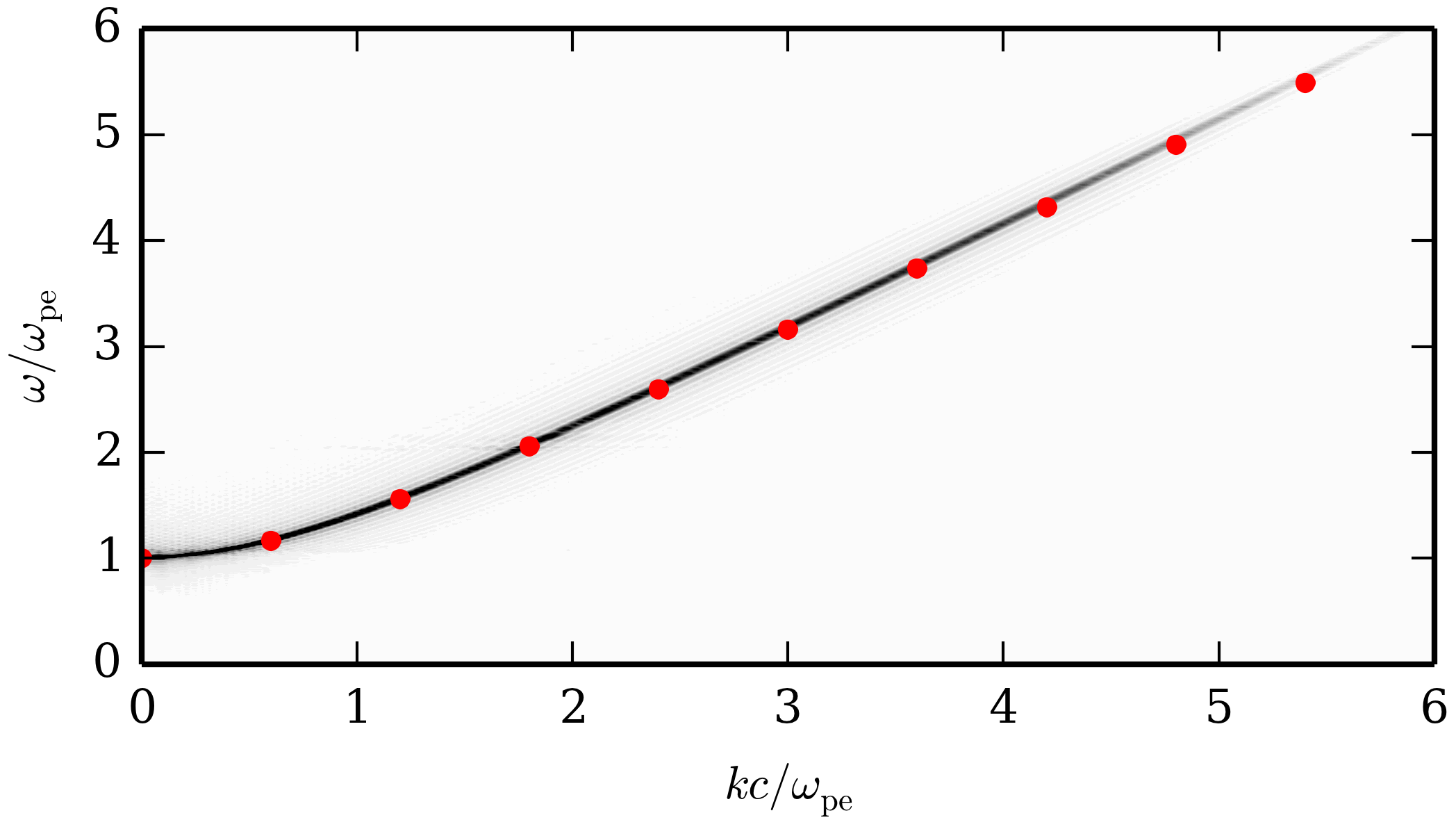}
}
\caption{Dispersion relation of a bulk plasma calculated using PIC simulations (axes normalized to the electron plasma frequency). Red dots indicate reference values obtained from equation \eqref{eq:cold_plasma_dispersion_relation}.}
\label{fig:plasma_dispersion_relation}
\end{figure}

Many theoretical studies of PPCs rely on the cold-plasma model as a simplifying assumption for the interaction of the electromagnetic field with the plasma. It is, therefore, instructive to initially validate this assumption using PIC. This approach is arguably more fundamental as it entails a kinetic description of a representative ensemble of particles. The cold-plasma model \cite{swanson_plasma_2003,lieberman_principles_2005} is derived from a one-dimensional force balance for electrons \cite{lieberman_principles_2005}
\begin{align*}
m_\textrm{e} \frac{d v_\textrm{e}}{dt} + m_\textrm{e} \nu_\textrm{m} v_\textrm{e} = - e E.
\end{align*}
Therein, $e$ is the elementary charge, $m_\textrm{e}$ is the electron mass, $v_\textrm{e}$ is the electron velocity, and $\nu_\textrm{m}$ is the electron-neutral collision frequency. A plasma dielectric constant is found as \cite{lieberman_principles_2005}
\begin{align*}
\varepsilon_\textrm{p}(\omega) = 1 - \frac{\omega_\textrm{pe}^2}{\omega (\omega - i \nu_\textrm{m})}
\end{align*}
from Ampere's law, using a time-harmonic approach $f(t) = \textrm{Re}\{ \tilde{f}(\omega) \exp (i \omega t) \}$ and assuming a stationary ion background. $\omega_\textrm{pe}=\sqrt{e^2 n_\textrm{e} / (\varepsilon_0 m_\textrm{e})}$ is the electron plasma frequency and $\varepsilon_0$ is the vacuum permittivity. This result closely corresponds to the well-known Drude model \cite{drude_zur_1900}. In summary, this complex, frequency-dependent relative permittivity is derived assuming a homogeneous, quasi-neutral, and infinite plasma. Its dispersion relation (a function of the angular frequency $\omega$ and the wave number $k$) in the collisionless case ($\nu_\textrm{m} \rightarrow 0~\textrm{s}^{-1}$) is \cite{lieberman_principles_2005}
\begin{align}
	D(\omega, k) = \omega^2 - \omega_\textrm{pe}^2 - (k c)^2 = 0,
	\label{eq:cold_plasma_dispersion_relation}
\end{align}
with the vacuum speed of light $c$.

In a one-dimensional frame, this result can be compared to PIC simulation results of a thick plasma slab (mimicking a bulk plasma). As concerns the boundaries, electrons and ions are fully reflected, and as such no bounding plasma sheath can establish. Initially a homogeneous, quasi-neutral plasma is imposed. Due to particle conservation and negligible energy coupling of the electromagnetic fields into the plasma this state approximately remains throughout. Consequently, a homogeneous, quasi-neutral plasma is simulated, which is finite in size as concerns the electromagnetic fields.

A spatio-temporal plot of the electric field propagation through the plasma is shown in figure \ref{fig:plasma_propagation}. Despite the finite size of the simulated plasma slab, reflections of the electromagnetic field at the boundaries  are small. As the proceeding analysis is conducted in the frequency-domain, these reflections ``assist'' to confine the electromagnetic field to the slab region for a longer time period. The longer the effective response of the electromagnetic field propagation through the plasma, the more accurate the obtained spectra. The corresponding simulation results analyzed in the two-dimensional frequency (Fourier) domain are displayed in figure \ref{fig:plasma_dispersion_relation} along with the analytic result (indicated by red dots) calculated from equation \eqref{eq:cold_plasma_dispersion_relation}. For the latter, a remarkable agreement is observed, indicating that the reflecting boundary condition for the particles mimic an infinitely extended plasma. Note that the signal of the system response decreases with increasing frequency. This is merely due to the finite pulse width of the excitation rather than physical damping of the signal. From the agreement of the two models it can be argued that the PIC model very well justifies the use of the cold-plasma model for homogeneous plasmas, despite the low pressure. This aspect is well reasoned with the observation that for an infinite plasma, the ratio of the collision mean free path and the system size is $\lambda_\textrm{c}/L_\textrm{s} \rightarrow 0$. Note that $\lambda_\textrm{c}$ is approximately related to the collision frequency $\nu_\textrm{m}$ via $\lambda_\textrm{c} = v / \nu_\textrm{m}$, given the particle velocity $v$. The above reasoning is substantially different regarding the temporal dynamics as the collision frequency governs the randomization of the particle trajectories and, consequently, the damping of electromagnetic field response. (This is also why a low pressure $p=1$ Pa was chosen in this study.)

\subsection{Plasma photonic crystal}
\begin{figure}[b!]
\centering
\resizebox{8cm}{!}{
\includegraphics{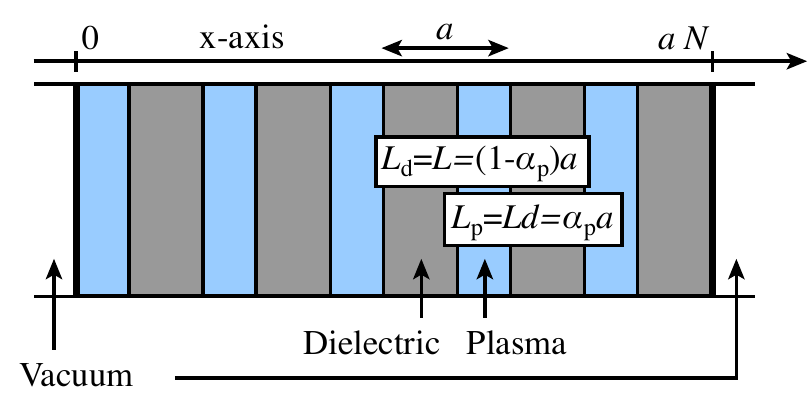}
}
\caption{Schematic of a finite size one-dimensional plasma-dielectric photonic crystal with lattice constant $a$, plasma thickness $L_\textrm{p}$ and thickness of the dielectric $L_\textrm{d}$. The x-axis denotes the region of finite extend.}
\label{fig:schematic}
\end{figure}

\begin{figure*}[t!]
\centering
\resizebox{16cm}{!}{
\begin{tabular}{ c c c }
  \includegraphics{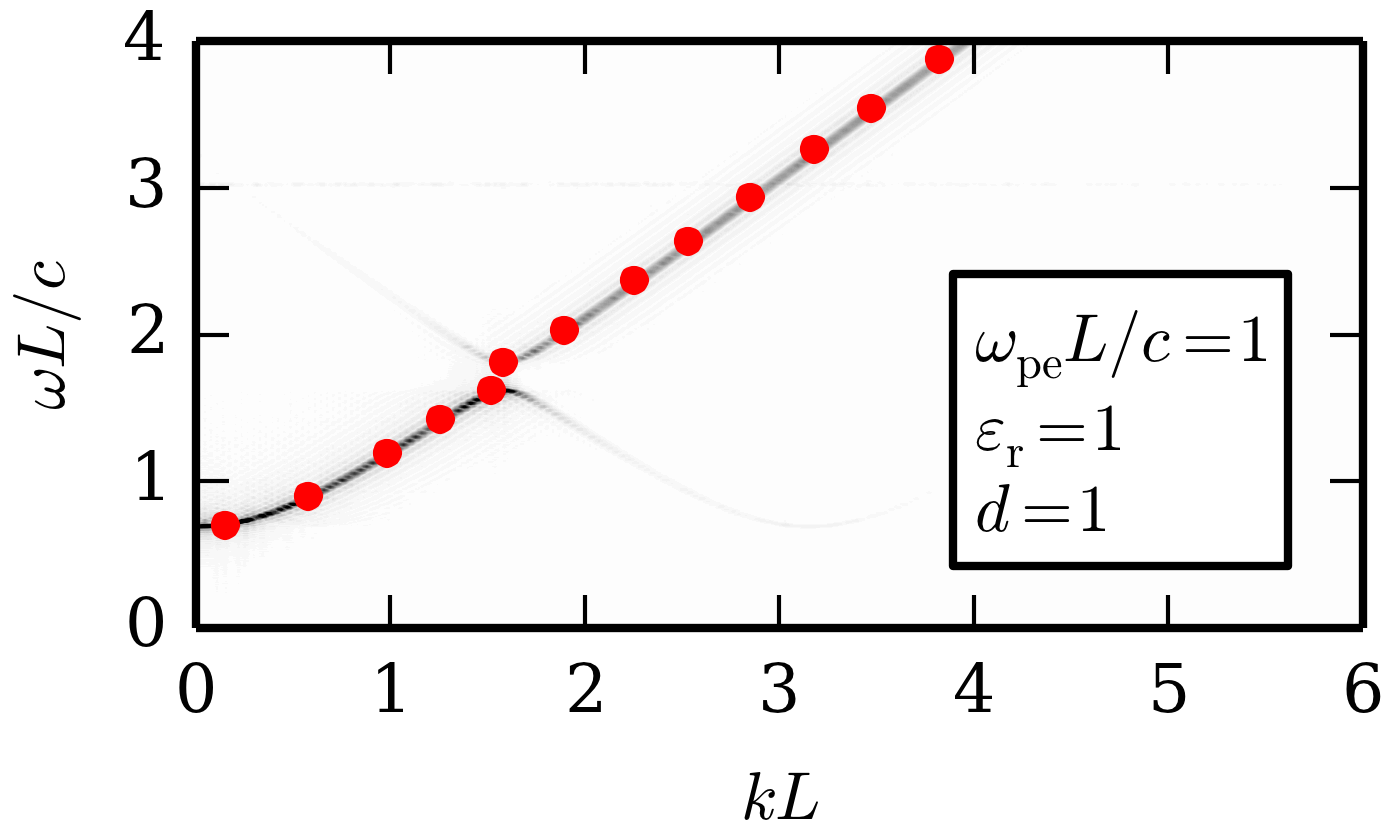} & 
  \includegraphics{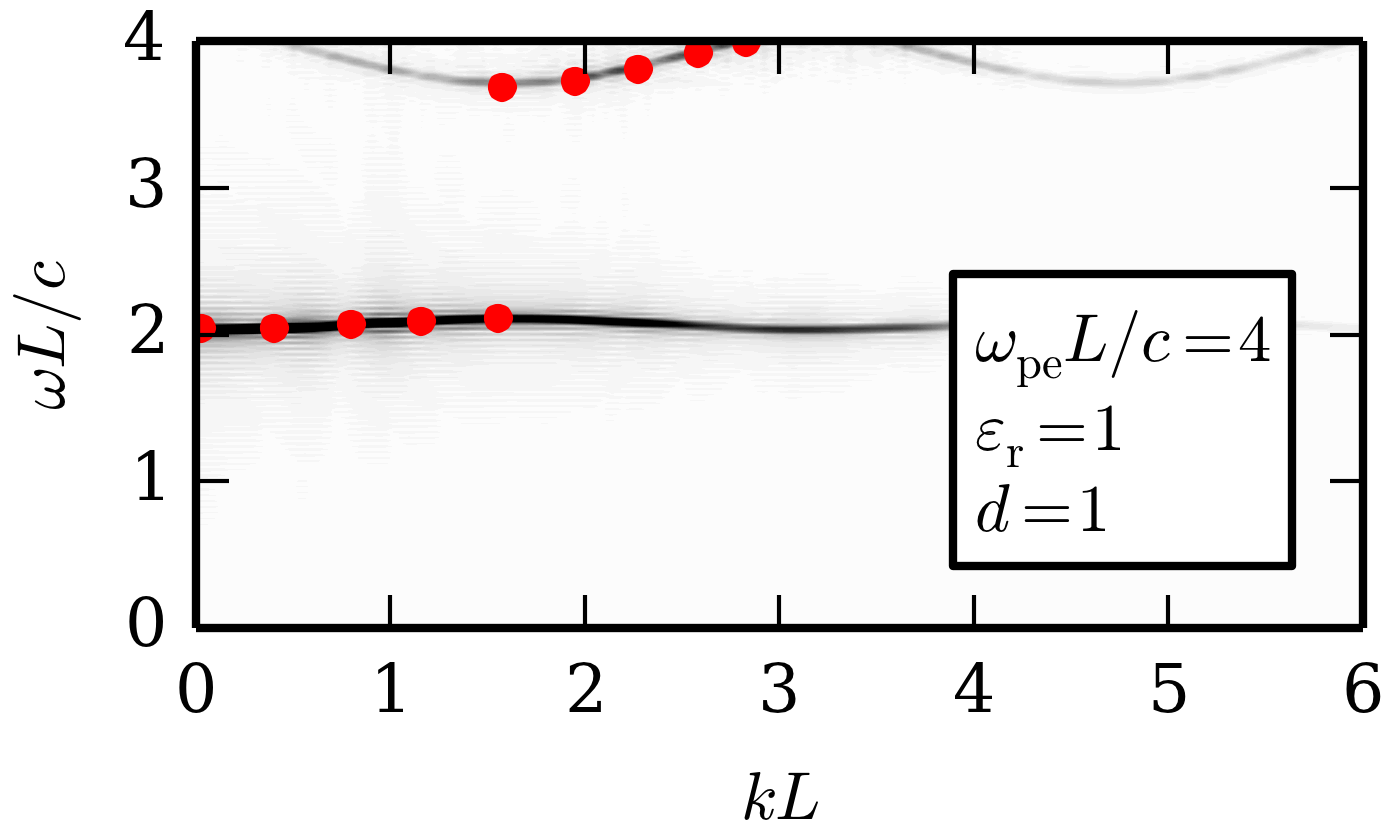} & 
  \includegraphics{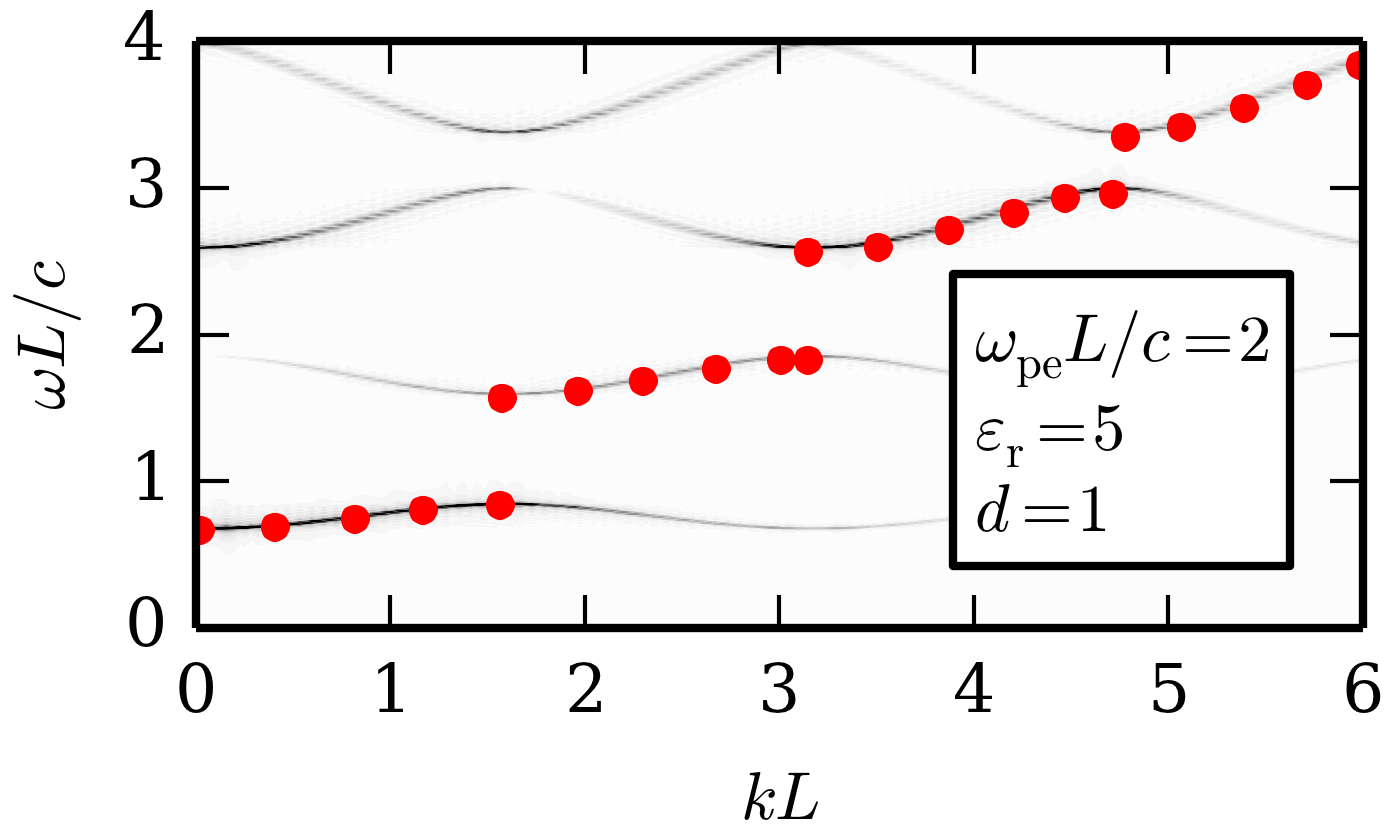} \\
  \includegraphics{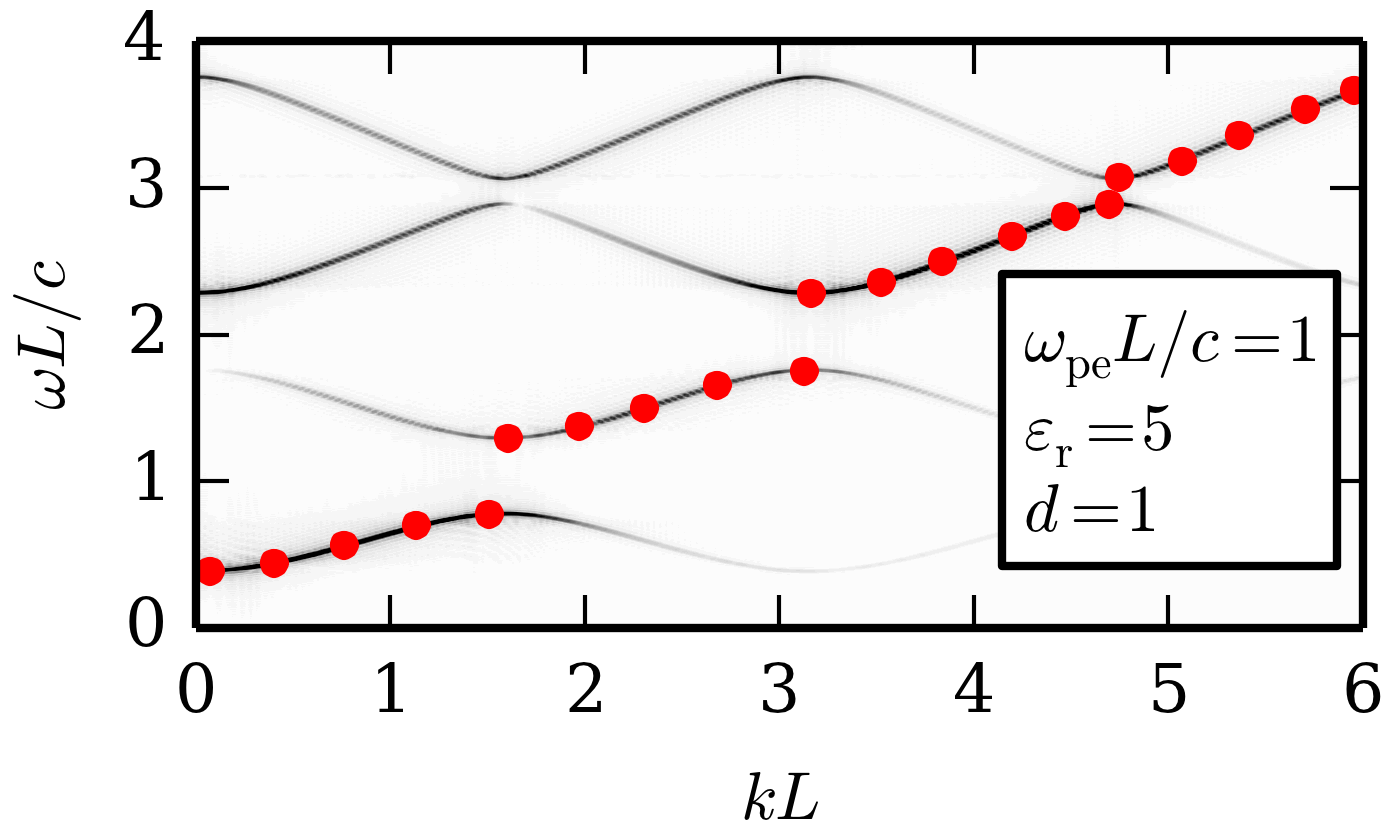} & 
  \includegraphics{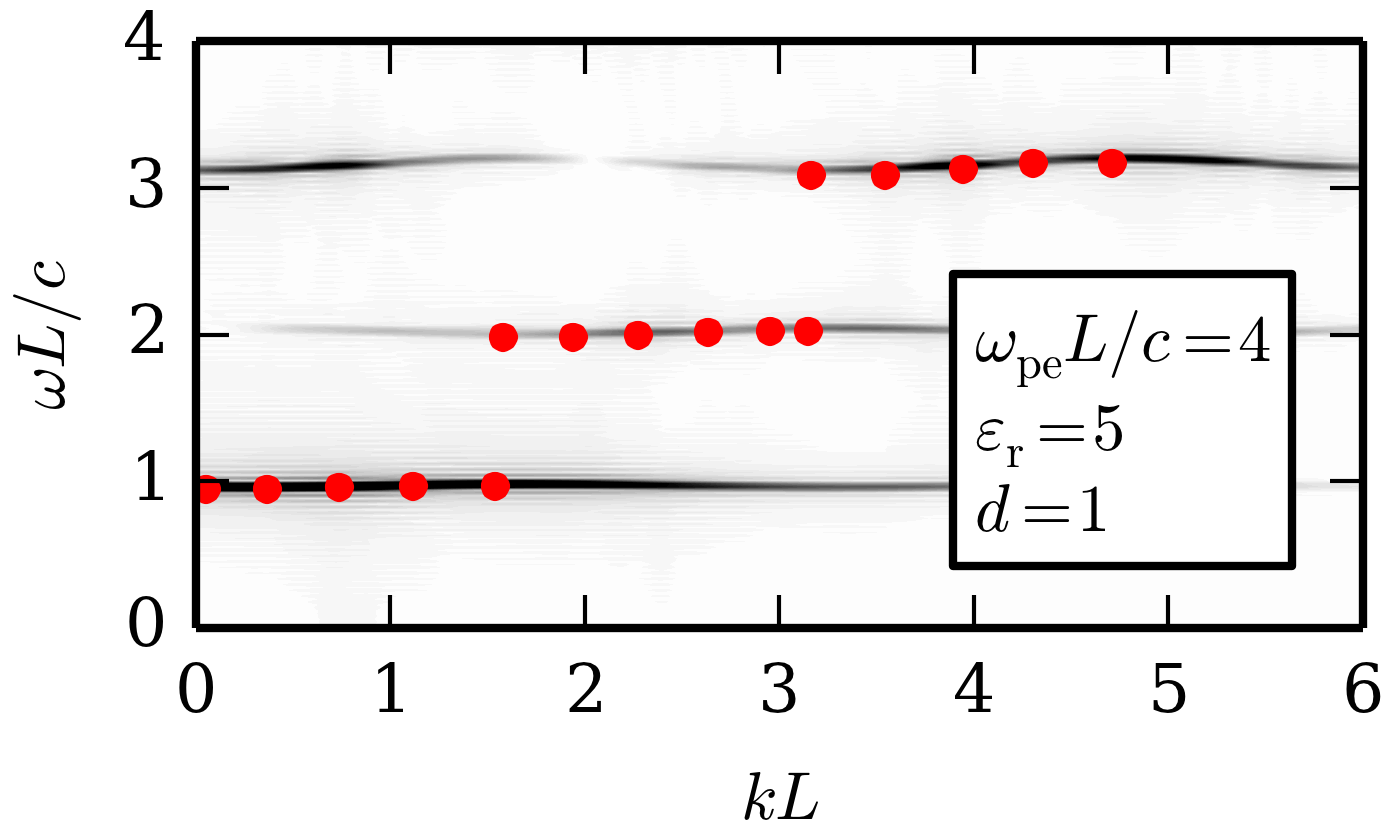} & 
  \includegraphics{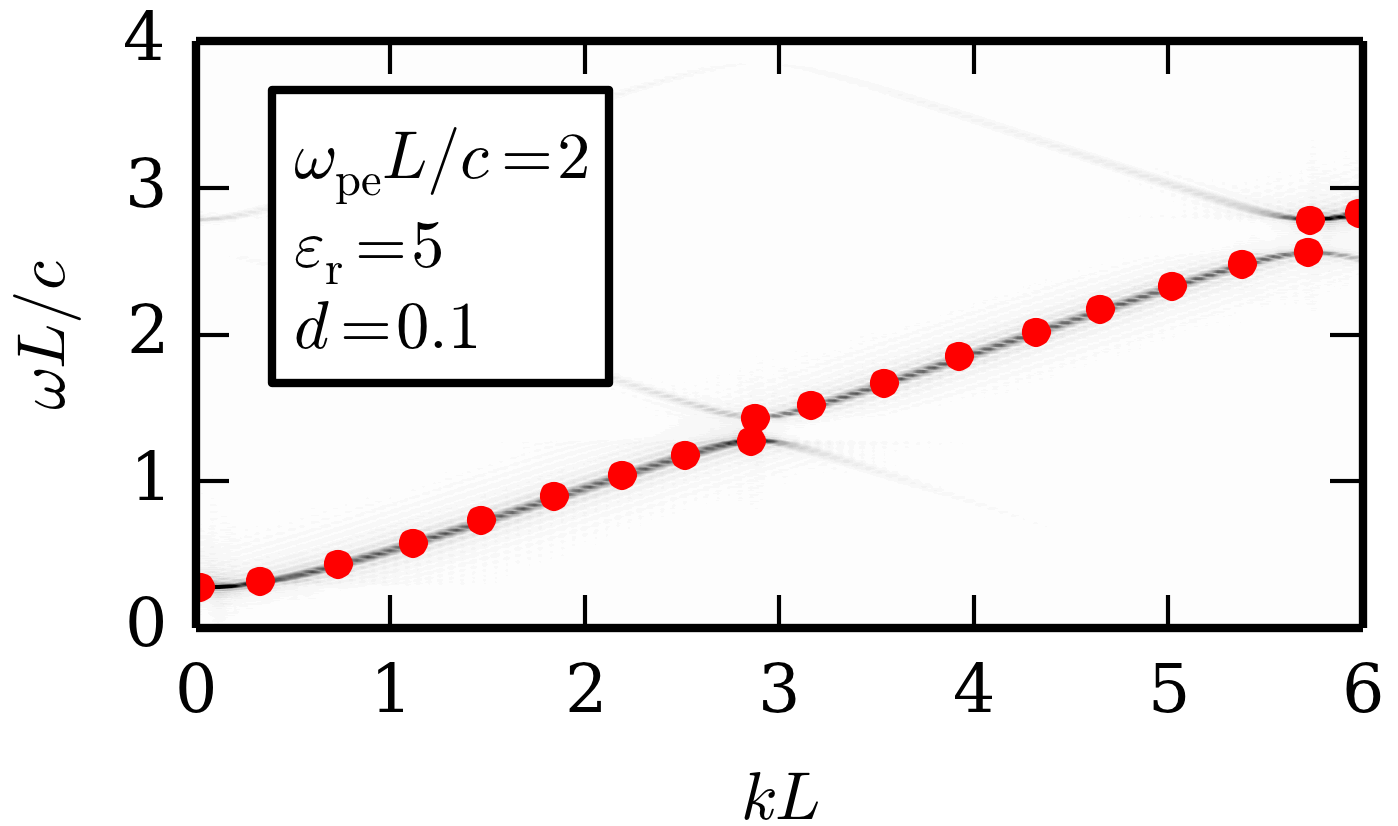} \\
\end{tabular}
}
\caption{Dispersion relations for different benchmark cases (cf. individual insets) calculated using PIC simulations. The red dots indicate values extracted from Hojo and Mase \cite{hojo_dispersion_2004} for comparison.}
\label{fig:dispersion_relation_benchmark}
\end{figure*}

A critical extension of our analysis concerns the application of the PIC model to the concept of plasma photonic crystals. This is particularly instructive because the cold-plasma model has been widely used to describe these scenarios. For validation, a geometry identical to the one considered by Hojo and Mase \cite{hojo_dispersion_2004} is analyzed. It is schematically depicted in figure \ref{fig:schematic}. Based on their notation, the one-dimensional photonic crystal is geometrically described in terms of the thickness of the dielectric $L_\textrm{d} = L$ and the plasma $L_\textrm{p} = Ld$, $L$ being a specified length. They provide a set of three selective parameters of a one-dimensional PPC, namely \{$\varepsilon_\textrm{r}$, $d$, $\omega_\textrm{pe} L/c$\}. Here $\varepsilon_\textrm{r}$ is the relative permittivity of the dielectric, $d$ is the relative thickness of the plasma (related to $L$), and $\omega_\textrm{pe} L/c$ is a plasma related parameter. Additionally, they provide equations for the dispersion relation of this setup based on these parameters. Most importantly, however, they base their theoretical results on the cold-plasma model. PIC simulations of this setup were performed using the model described previously with a correspondingly adapted geometric configuration. The plasma is divided into homogeneous layers of plasma partitioned by layers of dielectric (as shown in figure \ref{fig:schematic}). This configuration is identical to \cite{hojo_dispersion_2004}. In figure \ref{fig:dispersion_relation_benchmark}, PIC simulation results (in the Fourier domain) for the described PPC geometry are presented for the different cases discussed in \cite{hojo_dispersion_2004}. Reference values extracted from \cite{hojo_dispersion_2004} are again indicated by red dots and plotted for comparison. Both PPC dispersion relations are in remarkable agreement. Note that Hojo and Mase present the dispersion relation irrespective of the zone edges of the first irreducible Brillouin zone; this will be of interest later. An immanent difference lies in the way the results are obtained: The reference dispersion relations \cite{hojo_dispersion_2004} are calculated directly from the system properties. In contrast, the response obtained from PIC also depends on the mode coupling of the probe signal and the corresponding selective excitation. The visible outcome of the PIC results, therefore, largely depends on the coupling of the excitation to the respective modes available. In addition, the results are implicitly folded at the Brillouin zone edges by coupling into the neighboring cells (although only weakly). Therefore many more symmetry modes appear in the PIC results. It can be concluded that -- if particle-wall interactions with the layer boundaries are neglected and, hence, no plasma boundary sheaths are formed -- the cold-plasma model and the kinetic PIC approach yield the same results. In consequence, it seems appropriate to use the cold-plasma approach in these scenarios.

\subsection{Band diagram representation}
From the previous considerations it can be concluded that a dispersion relation based on the cold-plasma model is well justified for: (i) a homogeneous plasma slab and (ii) homogeneous layers of plasmas forming a PPC. Moreover, as indicated by the zone-folded dispersion diagrams obtained from the PIC model, results (ii) can be viewed in the context of Floquet's or Bloch's theory \cite{floquet_sur_1883,bloch_uber_1929,joannopoulos_photonic_2008} with a periodic description in the first irreducible Brillouin zone. A definition based on the lattice constant $a = L_\textrm{d} + L_\textrm{p}$ is suggested. The normalized angular frequency $\omega a / (2\pi c)$ and the normalized wave number $k a / (2\pi)$ follow and the analysis can be restricted to $-0.5 < ka / (2\pi) < 0.5$. In consequence, an alternative set of selective parameters can be used with \{$\varepsilon_\textrm{r}$, $\alpha_\textrm{p}$, $\omega_\textrm{pe} a/c$\}, which may be more favorable depending on interpretation. The parameters are the relative permittivity of the dielectric $\varepsilon_\textrm{r}$, the plasma plasma filling fraction of the PPC $\alpha_\textrm{p} = L_\textrm{p} / a$, and the plasma related parameter $\omega_\textrm{pe} a/c$ (now based on the lattice constant $a$). This definition of the variables and selective parameters implies that the scaling invariance property (cf.\ dielectric photonic crystals) fundamentally also apply to PPC -- including the plasma component. It follows that the relative plasma strength $\omega_\textrm{pe} a/c$ and the dielectric constant $\varepsilon_\textrm{r}$ of the intermediate dielectric layers in a PPC are closely connected to the two relative permittivities of a purely dielectric PC. It is, however, worthwhile to note that in a PPC there is of course a strong frequency dependence. The dielectric contrast largely differs in the frequency interval below the threshold of the electron plasma frequency $\omega a/c < \omega_\textrm{pe} a/c$ and above $\omega a/c > \omega_\textrm{pe} a/c$. Below the electron plasma frequency the plasma does not provide any propagating modes, thus making the PPC largely dissipative. A fact that may be undesired or intentionally utilized.

\begin{figure}[t!]
\centering
\resizebox{8cm}{!}{
\begin{tabular}{ c }
  \includegraphics{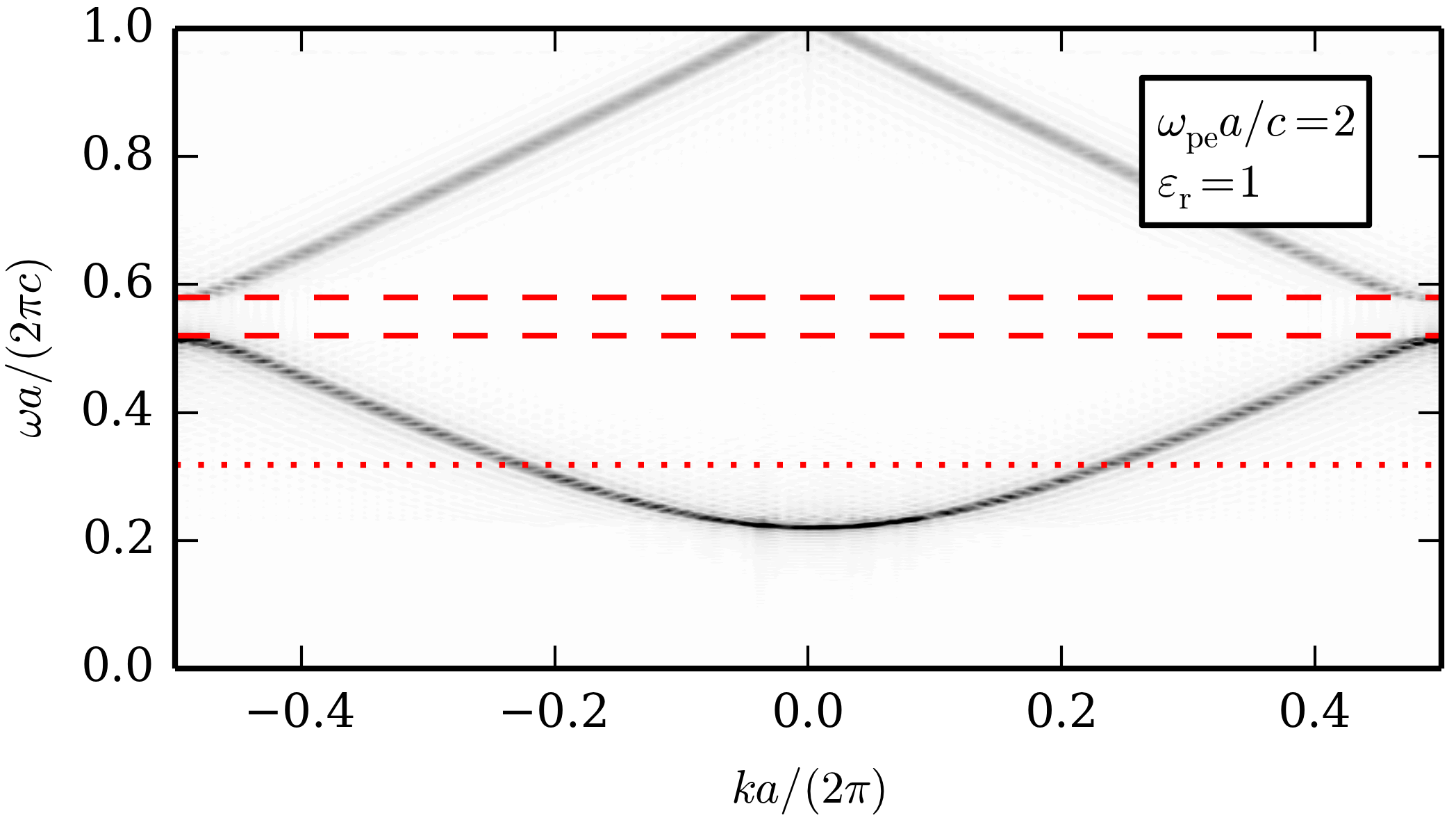} \\
  \includegraphics{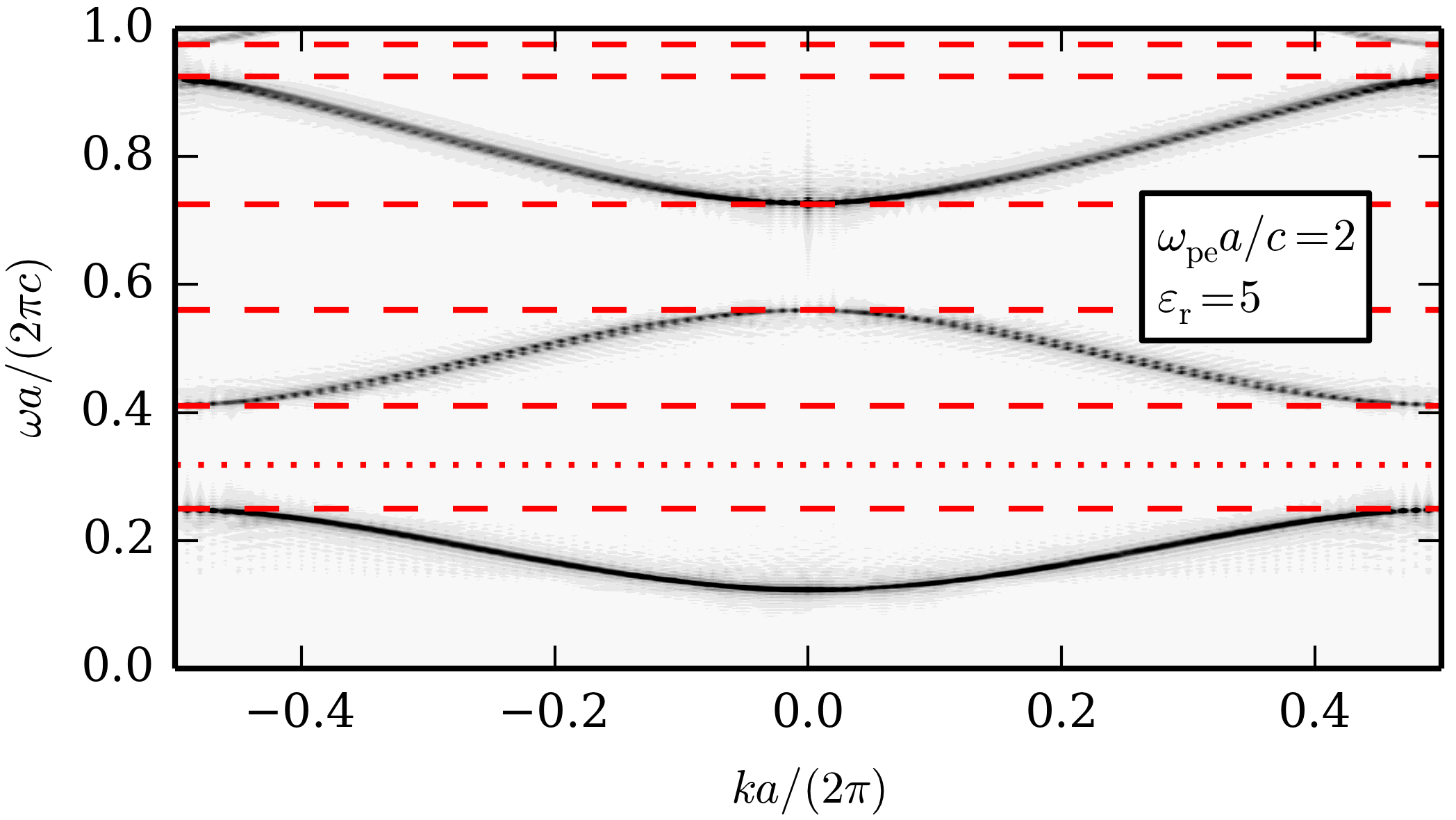} \\
  \includegraphics{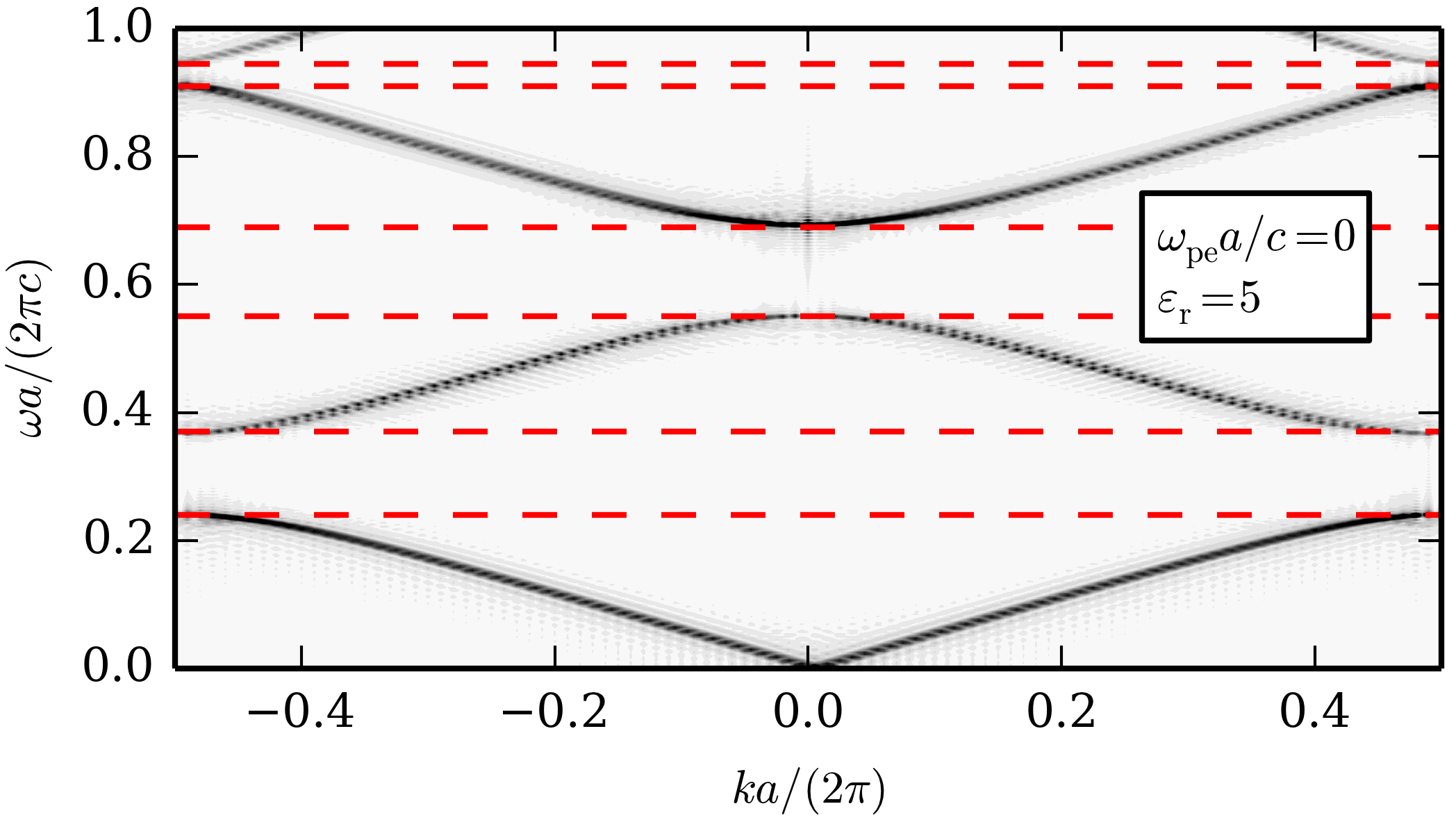}
\end{tabular}
}
\caption{Band diagram representations (first irreducible Brillouin zone) for a plasma photonic crystal with $\alpha_\textrm{p} = 0.5$ (that is $d = 1$) and different combinations of $\omega_\textrm{pe} a / c$ and $\varepsilon_\textrm{r}$ (top: plasma-air, center: plasma-dielectric, below: air-dielectric; see insets) obtained by PIC simulations. Bandgaps are marked by dashed red lines; $\omega_\textrm{pe}a/(2\pi c)$ is marked by dotted red lines.}
\label{fig:dispersion_zone_folded}
\end{figure}

Figure \ref{fig:dispersion_zone_folded} presents three exemplary cases in terms of photonic band diagrams for plasma-air (top), plasma-dielectric (center) and air-dielectric (below) PCs, respectively. Due to the full-wave approach of the PIC simulations, the photonic bands are -- to some extend -- implicitly folded at the zone edge. However not all modes are excited equally well (cf.\ figure \ref{fig:dispersion_relation_benchmark}). Therefore, in figure \ref{fig:dispersion_zone_folded} the modes were explicitly folded back into the first irreducible Brillouin zone at the zone edges. One can clearly distinguish the individual photonic bands as well as their separating bandgaps in this representation (cf.\ dashed red lines). In the case of air-separated plasma layers (figure \ref{fig:dispersion_zone_folded}, top) there is only a small difference compared to a homogeneous plasma slab. For the first bandgap that forms, only a small gap-midgap ratio of $\Delta \omega/\omega_\textrm{m} = 10.9\%$ is found ($\Delta \omega$ being the absolute size of the bandgap and $\omega_\textrm{c}$ being the midgap frequency). Moreover, the minimum of the lower band (dielectric band) at $ka/(2\pi) = 0$ is indeed quite close to the normalized electron plasma frequency at $\omega_\textrm{pe}a/(2\pi c) = 0.32$ (cf.\ dotted red line). This is due to (i) the low dielectric contrast of the PPC itself and (ii) the fact that the lowest dielectric constant of the PPC corresponds to vacuum. Following the (in this case possibly misleading) nomenclature of \cite{joannopoulos_photonic_2008}: the energy of the dielectric band (below the bandgap) is mostly concentrated in the plasma layers, while the energy of the air band (above the bandgap) resides mostly in the dielectric layers in between -- in this case air. The behavior is different for the plasma-dielectric case (figure \ref{fig:dispersion_zone_folded}, center): As expected, the higher dielectric constant of the intermediate layers pulls down all bands to lower frequencies. The gap-midgap ratio of the first bandgap thus significantly increases to $\Delta \omega/\omega_\textrm{m} = 48.5\%$ (a combination of an increased absolute bandgap $\Delta\omega$ and a decreased midgap frequency $\omega_\textrm{m}$). In this case, the lowest band lies significantly below the normalized electron plasma frequency. Consequently, propagating modes are allowed which are forbidden in a homogeneous plasma. Finally, in the case of a purely dielectric PC (figure \ref{fig:dispersion_zone_folded}, below) the situation for frequencies above $\omega a/(2\pi c) \approx 0.23$ is quite similar. The high dielectric constant also pulls down the bands to lower frequencies. Below the mentioned frequency, however, the difference between the plasma and the vacuum becomes apparent: Propagating modes are found for frequencies below the first bandgap down to  $\omega a/(2\pi c) \rightarrow 0$. This is not the case for the plasma-dielectric and plasma-air PCs, which are governed by the plasma cut-off frequency. In addition, the purely dielectric gap-midgap ratio of the first bandgap is slightly smaller with $\Delta \omega/\omega_\textrm{m} = 42.6\%$. With respect to potential applications in particular the first aspect may be important, as a distinct cut-off frequency (known for the cold-plasma dispersion relation) may be utilized to deliberately select the PC's properties.

The procedure of embedding the plasma properties into the scaling invariant theory of purely dielectric PCs can be used straightforwardly. The approach can be readily extended to two or three dimensions. With respect to the cases discussed in this work, one has to be careful about the directionality of the bandgap. The performed calculations are strictly one-dimensional -- in terms of the configuration as well as the analysis. Taking into account oblique incidence, it follows from the one-dimensional geometry that the PPC does not have a complete bandgap. However, depending on the dielectric constant omnidirectional reflection might still occur \cite{joannopoulos_photonic_2008}. As our simulation is at this point restricted to one dimension only, this is beyond the scope of this work.

\section{Conclusion}
To summarize this work, the plasma dispersion relation based on the cold-plasma model -- frequently used as a simplifying assumption -- has been verified by means of a kinetic approach. Moreover, by using PIC simulations the theoretical results of Hojo and Mase \cite{hojo_dispersion_2004} were verified and the assumption of a cold-plasma was validated for a multilayer geometry of homogeneous layers of plasma. The PIC model is well capable of describing complex particle interactions with the walls and in consequence the formation of plasma boundary sheaths, which may alter the results to some extend. This is, however, beyond the focus of this work, which is on an alternative representation of the presented results in the frame of PC theory. A corresponding set of alternative, selective parameters was suggested. Hence, a scaling invariant interpretation was provided is, which is valid as long as non-linear interactions (such as substantial heating of the plasma by the incident radiation) or the invalidity of the cold-plasma assumption can be neglected. The formulation can be readily applied to multiple dimension.

\section*{Acknowledgments}
The authors thank Peter Edenhofer from Ruhr University Bochum for valuable comments and fruitful discussions. This work is supported by the Deutsche Forschungsgemeinschaft (DFG) in the frame of Collaborative Research Centre TRR 87 (SFB TR-87). 

\bibliography{references}

\end{document}